# Frequency dependence of specific heat in supercooled liquid water and emergence of correlated dynamics

**Shinji Saito**[1,a)], **Iwao Ohmine**[1,b)] **and Biman Bagchi**[2,c)]

[1] Institute for Molecular Science, Myodaiji, Okazaki, Aichi, 444-8585, Japan, and
The Graduate University for Advanced Studies, Myodaiji, Okazaki, Aichi, 444-8585, Japan

[2] Indian Institute of Science, Bangalore, 560012, India

Molecular origin of the well-known specific heat anomaly in supercooled liquid water is investigated here by using extensive computer simulations and theoretical analyses. A rather sharp increase in the values of isobaric specific heat with lowering temperature and the weak temperature dependence of isochoric specific heat in the same range are reproduced in simulations. We calculated the spatiotemporal correlation among temperature fluctuations and examined the frequency dependent specific heat. The latter shows a rapid growth in the low frequency regime as temperature is cooled below 270 K. In order to understand the microscopic basis of this increase, we have performed a shell wise decomposition of contributions of distant molecules to the temperature fluctuations in a central molecule. This decomposition reveals the emergence, at low temperatures, of temporally slow, spatially long ranged large temperature fluctuations. The temperature fluctuation time correlation function (TFCF) can be fitted to a William-Watts stretched exponential form with the stretching parameter close to 0.6 at low temperatures, indicating highly non-exponential relaxation. Temperature dependence of the relaxation time of the correlation function can be fitted to Vogel-Fulcher-Tamermann expression which provides a quantitative measure of the fragility of the liquid. Interestingly, we find that the rapid growth in the relaxation time of TFCF with lowering temperature undergoes a sharp crossover from a markedly fragile state to a weakly fragile state around 220 K.

a) Electronic mail: shinji@ims.ac.jp
b) Electronic mail: ohmine@ims.ac.jp
c) Electronic mail: bbagchi@sscu.iisc.ernet.in

# I. INTRODUCTION

Water is the most abundant liquid on earth. It displays a large number of anomalous properties even at ambient condition, for example the presence of the maximum of density in liquid state.[1-3] The anomalies are enhanced in its supercooled liquid state. For example, the isobaric specific heat, $C_P$, increases with decreasing temperature. The increase in $C_P$, down to the lowest temperature of 235 K, has been measured by several groups.[4-7] The isothermal compressibility, $\kappa_T$, also shows an increase with decreasing temperature. Many decades ago Speedy and Angell proposed that the thermodynamic response functions diverge at 228 K.[8] However, it is impossible to experimentally examine the static structure and thermodynamic properties of the supercooled water directly, because crystallization intervenes through homogeneous nucleation at $T \sim 232$ K.

There is yet no agreement or consensus about what causes the thermodynamic anomalies in supercooled water, although several hypotheses have been proposed.[9-12] Many simulation studies seem to support the liquid-liquid critical transition (LLCP) scenario.[13-20] The results of the analyses of supercooled water in confined media, e.g. a micelle template mesoporous silica matrix, in which water does not crystalize, also support the LLCP scenario,[21-24] although there are also critical discussions on the behavior of water in the confined system.[25, 26] Recently, Limmer and Chandler have disputed the LLCP scenario[27] based on their simulations with mW potential that consists of both two- and three-body interactions.[28] However, Moore and Molinero have shown the evidence of a liquid-liquid transformation at ~202 K.[29] In addition to a lack of consensus on the scenario for the thermodynamic anomalies in the supercooled water, our understanding of molecular origin of the anomalies is still limited, as noted earlier.

In particular, the markedly anomalous temperature dependence of $C_P$, and the lack of it for the isochoric specific heat, $C_V$, has remained largely unexplained, although reproduced earlier in simulation studies.[20, 30-32] The $C_P$ is related to entropy fluctuations at constant pressure. These fluctuations themselves are related, at low temperature, to the transitions between inherent structures of the liquid, though the relationship is not simple. Large fluctuations always imply a flattening of free energy surface. Thus, the studies of the difference between $C_P$ and $C_V$ and of the temperature fluctuation time correlation functions (TFCFs) can help understanding some aspects of complex dynamics of water at low temperatures.

In the present manuscript, therefore, we concentrate on the elucidation of the molecular origin of the increase in $C_P$. It should be noted, here, that $C_V$ is not found to exhibit any such an increase as $C_P$, though $C_V$ has not been measured directly in supercooled state. We examined the effects of temporally correlated dynamics present in the temperature fluctuation in terms of the complex specific heats for $C_P$ and $C_V$, by using long molecular dynamics simulations at several temperatures, starting from 300 K down to 200 K.

The emergence of spatio-temporal correlations has also been explored by calculating the frequency and wave-vector dependent temperature fluctuations and also the shell-wise decomposition of the contributions of temperature fluctuations. We find the emergence of the correlated hydrogen bond network (HBN) dynamics, which temporally extends up to several nano-seconds and spatially up to the sixth hydration shell, thus consisting of more than 340 molecules at ~220 K, the maximum temperature of $C_P$. We showed that the temperature dependence of the relaxation times of temperature fluctuation is well fit with a Vogel-Fulcher-Tammann (VFT) law for both above and below the crossover temperature 220 K.

## II. THEORY

We elucidate the detailed dynamics of the temperature fluctuation by using the complex specific heat.[33-38] Here, we calculated the complex specific heat based on the derivation by Grest and Nagel.[33] The specific heat is given by

$$C = \left( \frac{1}{N_f} - \frac{N\langle \delta T^2 \rangle}{T^2} \right)^{-1}, \tag{1}$$

where $N_f$ is the numbers of degrees of freedom per molecule and $T$ and $\delta T$ are temperature and the temperature fluctuation calculated from the instantaneous kinetic energy of the system with $N$ molecules, respectively. By introducing the following time correlation function given by

$$K(t) = \frac{N \langle \delta T(0) \delta T(t) \rangle}{T^2}, \tag{2}$$

where now the second term in Eq. (1) is expressed by $K(0)$. By using the Fourier-Laplace transform of the time derivative of $K(t)$, $\dot{K}(t)$, the specific heat is expressed as

$$C = \left( \frac{1}{N_f} + \hat{\dot{K}}'(0) \right)^{-1}. \tag{3}$$

By generalizing the static specific heat given in Eq. (3) to the frequency dependent specific heat, we have the following equation,

$$\hat{C}(\omega) = \left( \frac{1}{N_f} + \hat{\dot{K}}(\omega) \right)^{-1} \tag{4}$$

by using the Fourier-Laplace transform of $\dot{K}(t)$. In order to compute complex isobaric specific heat free from the artifacts of thermostat and barostat, we carried out the MD simulations with the *NVE* conditions with the density determined in the *NPT* simulation at each temperature.

We define the time dependent temperature of a given shell in the following way. Since the temperature is defined as the average of the temperature of individual molecules, the instantaneous temperature of the system can be given as an average of the instantaneous temperature of all the shells of a molecule, e.g. molecule 1,

$$T(t) = \frac{1}{N} \sum_{n}^{All\ shells} T_1^{(n)}(t) . \tag{5}$$

The temperature fluctuation of the system can also be decomposed as

$$\langle \delta T(0) \delta T(t) \rangle = \frac{1}{N} \sum_{n,m} \sum_{i}^{N} \langle \delta T_i^{(n)}(0) \delta T_i^{(m)}(t) \rangle = \sum_{n,m} C^{(n,m)}(t) , \tag{6}$$

$$C^{(n,m)}(t) \equiv \frac{1}{N} \sum_{i}^{N} \langle \delta T_i^{(n)}(0) \delta T_i^{(m)}(t) \rangle \tag{7}$$

where $\delta T_i^{(n)}$ is the temperature fluctuation of *n*th shell of molecule *i*. The shell-wise molecular contribution of the temperature fluctuation of the HBN dynamics is calculated as $\hat{C}^{(n,n)}(\omega_{HB}) / N^{(n)}$, where $\omega_{HB}$ is the characteristic frequency of the HBN dynamics and $N^{(n)}$ is the number of water molecules in the *n*th shell. In the present analysis, shells are defined as the minima in the radial distribution function located at 3.3, 5.7, 7.5, 9.7, 11.8, and 13.7 Å for the first to sixth shell.

We carried out MD simulations with both *NVE* and *NPT* conditions. The Nóse-Hoover thermostat and Berendsen barostat were used in the simulations under the *NPT* conditions.[39] The TIP4P/2005 model potential was used for the water molecules.[40] The system size is 1000 water molecules. The periodic boundary condition was employed and the long-range electric interactions were calculated by using the Ewald sum. We performed two or three independent MD simulations at 300, 270, 250, 240, 230, 225, 220, 215, 205, 200 K and 190 K, e.g., three MD simulations, one 250 ns and two 150 ns simulations, at 220 K. The pressure is 1 atm in the simulations with *NPT* condition. MD simulations with 8000 molecules were also carried out at 300, 250, 220, and 205 K, to examine the system-size dependence. No significant system-size dependence was found.

## III. TEMPERATURE DEPENDENCE OF SPECIFIC HEAT: ORIGIN OF DIFFERENCE BETWEEN $C_P$ AND $C_V$

The temperature dependences of $C_P$ and $C_V$ calculated from MD simulations are presented in Fig. 1(a). There is no significant difference between $C_P$ and $C_V$ around room temperature. Both $C_P$ and $C_V$ slightly increase with the decrease in temperature from 300 K to 230 K. Below 230 K, $C_V$ decreases, whereas $C_P$ suddenly increases and reaches its maximum at ~220 K. Below 220 K, $C_P$ rapidly falls and it coincides with $C_V$ again below 200 K. The present results are in agreement with the previous calculated results.[20, 30-32] In addition, the calculated $C_P$ reproduces the rapid increase of $C_P$ experimentally observed.[4-7]

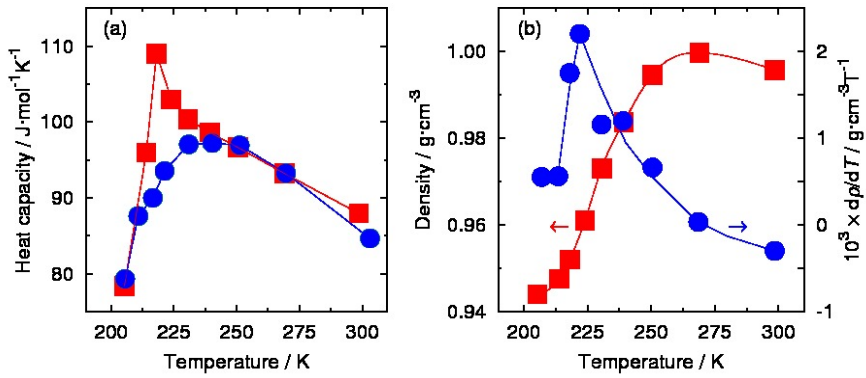

**FIG. 1**: Temperature dependences of (a) $C_P$ (red line) and $C_V$ (blue line) and (b) the density (red line) and temperature derivative of density (blue line).

The temperature dependence of the density calculated from the MD simulations under constant pressure condition is shown in Fig. 1(b). The calculated density increases with decreasing temperature from room temperature to ~270 K and then it decreases with further decreasing temperature. The density changes from ~0.97 to ~0.94 g/cm$^3$ in the temperature range from 230 K to 200 K is significantly the same range where the large difference between $C_P$ and $C_V$ is seen. The temperature derivative of density, $d\rho/dT$ is also shown in Fig. 1(b). The $d\rho/dT$ increases with decreasing temperature from room temperature and reaches the maximum at ~220 K. The temperature of maximum of $C_P$ almost coincides with the temperature of maximum of $d\rho/dT$. The consistency of the temperatures of the maximum of $C_P$ and $d\rho/dT$ suggests the relationship between change in $C_P$ and the change in dynamics caused by decrease in density.

## IV. COMPLEX SPECIFIC HEAT: EMERGENCE OF SLOW DYNAMICS

In order to understand the molecular motions contributing to the specific heats, we have calculated the real and imaginary parts of the complex specific heats shown in Fig. 2. The value at $\omega=0$ in the real part is the static specific heat. In order to understand results in Fig. 2, we have calculated TFCF, shown in Fig. 3. The peaks at ~100, ~400, and ~1000 cm$^{-1}$ in the spectra of the imaginary part correspond to the intermolecular motions, i.e. O-O-O band, O-O stretch, and rotation, exhibiting the fast oscillatory motions in the TFCF in Fig. 3. It is noted that the frequencies of oscillatory motions in the complex specific heat are observed at double the frequencies in spectra probing oscillatory amplitudes, e.g. IR, Raman, and neutron scattering measurements.[41] In addition to the peaks arising from the intermolecular motions, another peak is observed below 10 cm$^{-1}$. This peak is attributed to the HBN dynamics observed as the slow component in the TFCF presented in Fig. 3.

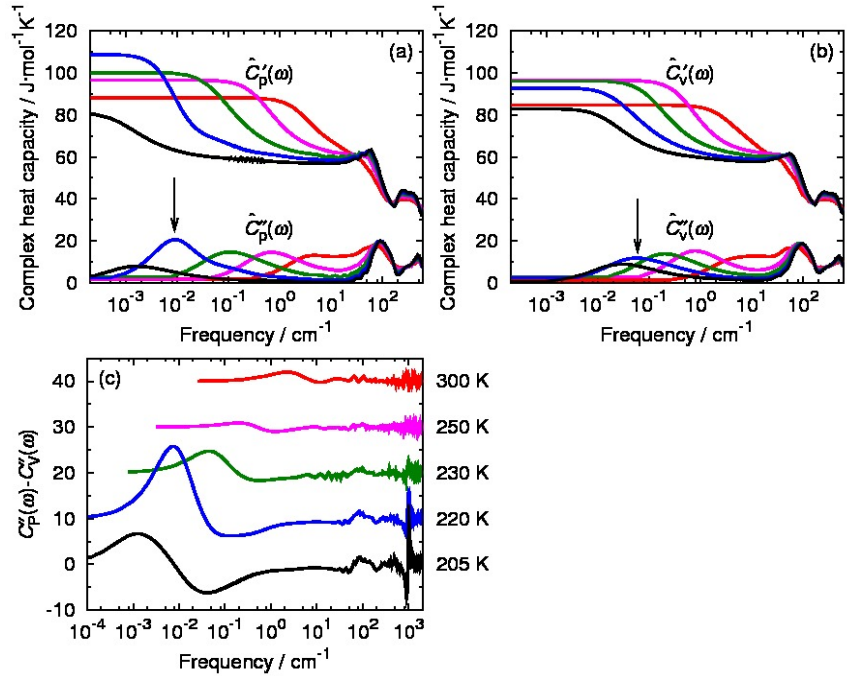

**FIG. 2**: Temperature dependences of the complex specific heats at densities determined (a) under constant pressure condition and (b) at constant volume with $\rho$ = 1.0 g/cm$^3$, and (c) difference between $\hat{C}_P''(\omega)$ and $\hat{C}_V''(\omega)$, respectively. Red, pink, green, blue, and black lines correspond to the results at 300, 250, 230, 220, and 205 K, respectively. Arrows show the peak position of HBN dynamics at 220K.

Above 250 K, $\tilde{C}_P(\omega)$ and $\tilde{C}_V(\omega)$ are similar to each other in a whole range of frequencies. Below 230 K, however, we find two noteworthy differences between $C_P$ and $C_V$. One is the difference between the time scale of the HBN dynamics between two conditions, e.g. the peak frequency of the HBN dynamics (~0.01 cm$^{-1}$) under constant pressure condition at 220 K is about six times lower than that at $\rho = 1.0$ g/cm$^3$. Another is the difference in intensity, i.e. the density of states, of the HBN dynamics contributing to the temperature fluctuation. The peak intensity of the HBN dynamics in $\tilde{C}_V''(\omega)$ gradually decreases due to the progress of the freezing of motions in supercooled water with decreasing temperature, whereas that in $\tilde{C}_P''(\omega)$ increases and shows a maximum at ~220 K under constant pressure condition. It is noted that the larger value of $\hat{C}_P''(\omega_{HB})$ than $\hat{C}_V''(\omega_{HB})$ shows that HB structural changes take place more efficiently under constant pressure condition, whereas they hardly occur under constant volume condition because of the lack of free volume.

The present analysis establishes that the origin of the well-known difference between $C_P$ and $C_V$ lies in these two differences between HBN dynamics under the two conditions. Furthermore, the present results demonstrate the difference in energy landscape and resultant dynamics between two conditions; liquid structures connected by low energy barriers can be reached by local volume fluctuations under constant pressure condition and, thus, structural changes proceed at a low density because of the presence of free volume under constant pressure condition.

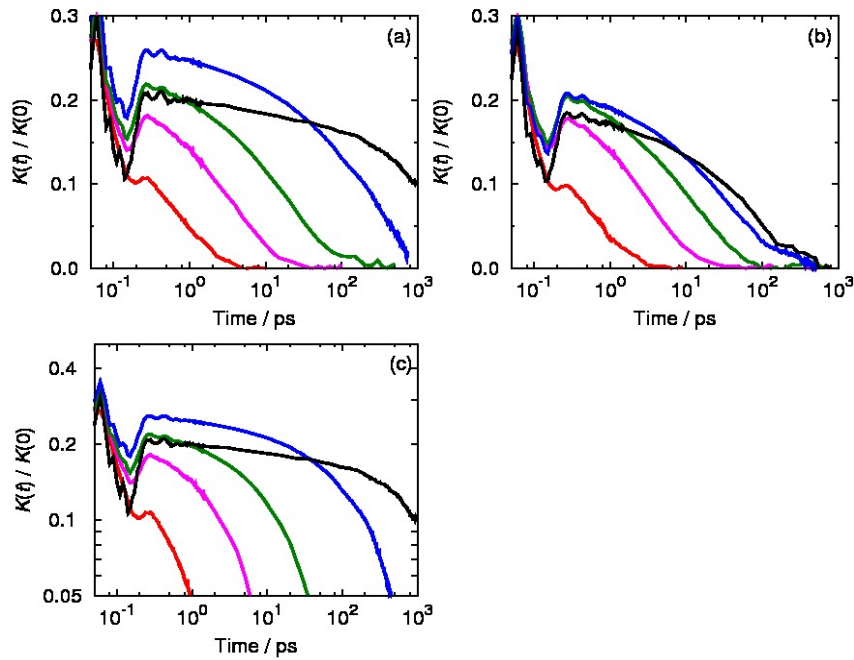

**FIG. 3**: Temperature dependences of the TFCFs (a) at densities determined under constant pressure condition and (b) at constant volume with $\rho = 1.0$ g/cm$^3$. Red, pink, green, blue, and black lines correspond to the results at 300, 250, 230, 220, and 205 K, respectively. (c) Log-log plots of the TFCFs at 220 (blue), 205 (black), and 200 (gray) K, respectively.

A static susceptibility is expressed by using the imaginary part of the corresponding complex susceptibility based on the Kramers-Kronig relation,

$$\chi = \frac{2}{\pi} \int_0^\infty \frac{\hat{\chi}''(\omega)}{\omega} d\omega . \tag{8}$$

By using the above relation, the difference between $C_P$ and $C_V$, $\Delta C$, is related to the difference between the imaginary part of the complex specific heats, $\hat{C}_P''(\omega)$ and $\hat{C}_V''(\omega)$,

$$\Delta C = \frac{2}{\pi} \int_0^\infty \frac{\hat{C}_P''(\omega) - \hat{C}_V''(\omega)}{\omega} d\omega . \tag{9}$$

The difference between $\hat{C}_P''(\omega)$ and $\hat{C}_V''(\omega)$ corresponding to the numerator in Eq. (8) is shown in Fig. 2(c). Two differences are found below ~10 cm$^{-1}$ and above ~100 cm$^{-1}$, in particular, at low temperatures. The difference from ~800 to ~1000 cm$^{-1}$ arises from the increase in the gap between the translational and rotational motions, due to the growth of correlated structure with decreasing temperature and density. The small difference from 100 to 300 cm$^{-1}$ is attributed to the difference between the intermolecular translational motions, O-O-O bend and O-O stretch. The difference below ~ 10 cm$^{-1}$ arises from the slow-down of the HBN dynamics under constant pressure condition compared with that under constant volume condition. Table I shows the contributions of the HBN dynamics and intermolecular translational and rotational motions to the difference between $C_P$ and $C_V$ calculated by the Kramers-Kronig relation. Since both the frequency and the intensity of $\hat{C}''(\omega)$ contribute to the specific heat, the contribution of the intermolecular motions with higher frequency than the HBN dynamics to $\Delta C$ is markedly reduced and, thus, $\Delta C$ indeed arises from the difference between the HBN dynamics under the two conditions.

**TABLE I**: Contributions of the HBN dynamics and the intermolecular translational and rotational motions to the difference between $C_P$ and $C_V$.

| Temperature (K) | $\Delta C^a$ (J/(mol K)) | $\Delta C_{KK}^b$ (J/(mol K)) | $\Delta C_{KK}^{HB}$ [c] (J/(mol K)) | $\Delta C_{KK}^{tra}$ [d] (J/(mol K)) | $\Delta C_{KK}^{rot}$ [e] (J/(mol K)) |
|---|---|---|---|---|---|
| 300 | 3.57 | 3.55 | 2.85 | 0.58 | 0.12 |
| 250 | 0.01 | 0.04 | 0.16 | -0.18 | 0.06 |
| 230 | 3.89 | 3.89 | 4.35 | -0.56 | 0.09 |
| 220 | 15.91 | 15.82 | 16.34 | -0.67 | 0.15 |
| 205 | -1.83 | -1.91 | -1.02 | -1.07 | 0.18 |

a: Difference between $C_P$ and $C_V$ calculated from molecular dynamics simulations.

b: Difference between $C_P$ and $C_V$ calculated by using the Kramers-Kronig relation.

c: Contribution of HBN dynamics defined as the motion whose frequency is less than 10 cm$^{-1}$.

d: Contribution of the intermolecular translational motion defined as the motion whose frequency is in between 10 and 700 cm$^{-1}$.

e: Contribution of the intermolecular rotational motion to defined as the motion whose frequency is in more than 700 cm$^{-1}$.

The slow component is expressed as a stretched exponential function at low temperatures,

$$C_{\text{slow}}(t) \sim a\exp[-(t/\tau_{\text{HB}})^\beta]. \tag{10}$$

The stretching parameter, $\beta$, can be related to heterogeneity in the system, i.e. the distribution of the relaxation times. As seen in Table II, the stretching parameter decreases with decreasing temperature up to 220 K at which it shows the minimum and then slightly increases with further decreasing temperature. This temperature dependence indeed indicates the increase in heterogeneity arising from the growth of tetrahedral structure in the system at ~ 220 K, and coincides with the large density fluctuation found around this temperature. It is also known that the stretching parameter is related to the fragility of the system.[42, 43] Thus, Table II also shows that the system is a markedly fragile above 220 K and becomes less fragile below the temperature.

**TABLE II**: Temperature dependences of stretching parameter, $\beta$, and fragility index, $D$, under constant pressure condition.

| Temperature (K) | 300 | 270 | 250 | 230 | 220 | 205 | 200 |
|---|---|---|---|---|---|---|---|
| $\beta$ | 1.0 | 0.96 | 0.70 | 0.60 | 0.56 | 0.60 | 0.60 |
| $D$ | 0.59 | 0.59 | 0.59 | 0.59 | 0.59 | | |
| | | | | | 1.25 | 1.25 | 1.25 |

Log-log plots of the TFCFs at 220, 205, and 200 K are presented in Fig. 3(c). A power law-type decay emerges below 220 K under the condition of constant pressure, though it can still be expressed by a stretched exponential function as shown above. In particular, the TFCF decreases in a linear fashion for about four decades of time (that is, for tens of ns duration) at 200 K. It is noted that the power-law decay is not found, at least above 205 K, under constant volume condition.

As seen in Fig. 3, a characteristic dip separates the fast and slow components. The location of the dip is insensitive to temperature, whereas its magnitude is strongly sensitive, increasing with lowering temperature. In addition, below 230 K, the magnitude of the dip under constant pressure condition is larger than that under constant volume condition. The magnitude reflects the rigidity of the liquid. Thus, the difference in the magnitude of dip between two conditions at low temperatures is related to the difference in their structures, i.e., the growth of long-range correlated structure.

We examined the temperature dependences of liquid structure. Figure S1a shows the temperature dependence of the distribution of volume of Voronoi polyhedra under the two conditions.[65] There is no significant difference between two conditions above 230 K. In contrast, a remarkable difference is found below 230 K; the distribution gradually shifts to a larger value under constant pressure condition with decreasing temperature, whereas almost no temperature dependence under constant volume condition. The present result shows that the distributions are unimodal at all the temperatures. This result is consistent with the continuous changes in $C_P$ and $d\rho/dT$ at 220 K.

The temperature dependence of the tetrahedrality[44] of water molecules is presented in Fig. S1b.[65] The distribution of tetrahedrality has a main peak is located at ~0.8 at 300 K. As in the volume of Voronoi polyhedra, the difference between the distributions of tetrahedrality under the two conditions is small above 230 K, whereas a remarkable difference is found below 230 K due to the growth of tetrahedral structure under constant pressure condition.[45]

A similar difference between the two conditions is found in the static structure factor of the center of mass of molecules, $S(k)$ (Fig. S1c of the supplementary material).[65] The $S(k)$ shows a doublet structure with peaks at ~2 and ~3 Å$^{-1}$. As experimentally known,[46] the low-$k$ peak at ~2 Å$^{-1}$ is sensitive to temperature and density and it shifts to a lower $k$ value with decreasing temperature below 230 K under constant pressure condition, whereas no clear temperature dependence of the low-$k$ peak is found under constant volume condition. All the above structural analyses demonstrate the growth of tetrahedral structure due to the increase in free volume per molecule with decreasing temperature and density under constant pressure condition, whereas the growth of tetrahedral structure is inhibited even at low temperatures under constant volume condition at $\rho = 1$ g/cm$^3$.

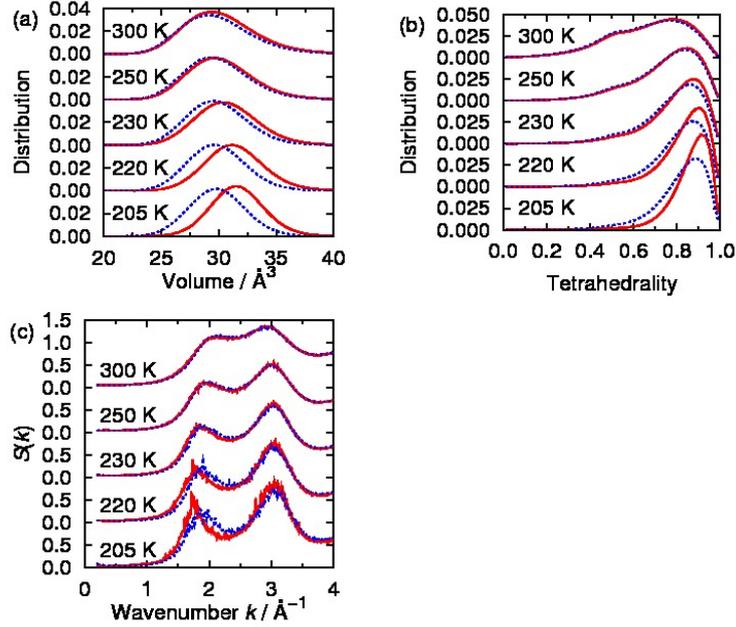

**Supplemental Figure 1**: Temperature dependences of the distributions of (a) volumes of Voronoipolyhedra, (b) tetrahedralities, and (c) static structure factors under constant pressure (red lines) and volume (blue lines) conditions.

## V. SPATIAL CORRELATIONS OF TEMPERATURE FLUCTUATIONS: EMERGENCE OF LONG LENGTH SCALE

A microscopic view of $\hat{C}_\mathrm{P}(\omega)$ can be obtained from wave number and frequency dependent temperature fluctuation, $|\hat{K}(k,\omega)|^2/S(k)$, scaled by the static structure of the center of mass of molecule, that provides a powerful tool to study coupling between length and time scales.[47] A shell-wise decomposition of contributions to the temperature fluctuation at the characteristic frequency of the HBN dynamics, as well as $|\hat{K}(k,\omega)|^2/S(k)$, are presented in Fig. 4.

As shown in Fig. 4(a), at 300 K, the main ($\omega$, $k$) peak of the HBN dynamics is found at (~2 cm$^{-1}$, ~1.7 Å$^{-1}$) that overlaps with the intermolecular vibration. A quantitative analysis of shell-wise contribution of temperature fluctuation reveals that the correlation within the first hydration shell is predominant at 300 K. With decreasing temperature, HBN dynamics are seen to decouple from the intermolecular vibration dynamics (Fig. 4(b)). The contributions from the shells further than the first shell grow and intermediate-range correlations become predominant starting from 240 K. With further decreasing temperature, the HBN dynamics contributing to the specific heat is well separated spatially and temporally from the intermolecular vibrations; correlated HBN dynamics are evident at $(\omega, k) = (\sim 0.01 \text{ cm}^{-1}, \sim 1.6 \text{ Å}^{-1})$ and the HBN dynamics has a tail from the low-$k$ peak of the doublet in $S(k)$ to still lower $k$ values (Fig. 4(c)). The correlated region extends up to the sixth hydration shell, which consists of ~350 water molecules (Fig. 4(d)). The result in Fig. 4(d) indicates the significant growth of correlation length towards with decreasing temperature up to 220-230 K, which is caused by the growth of tetrahedral structure due to the decrease in density with decreasing temperature.

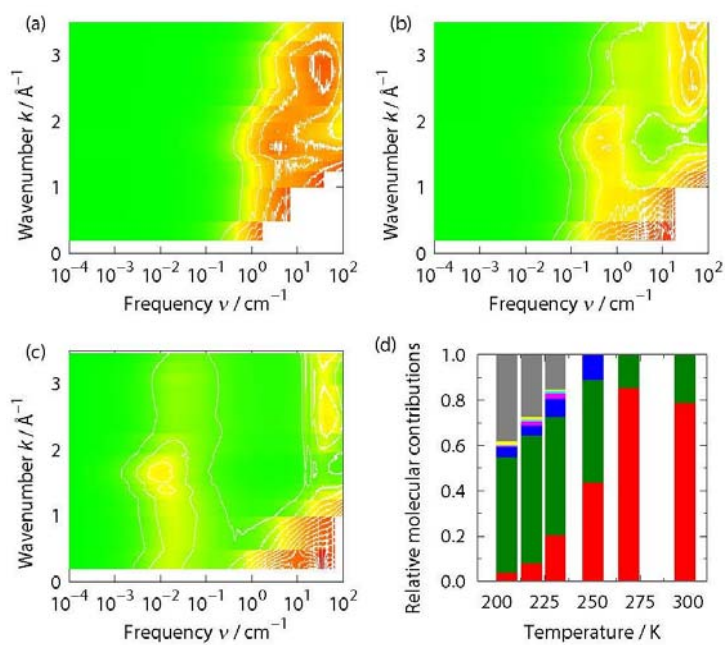

**FIG. 4**: Structure factor scaled frequency and wavenumber dependent temperature fluctuation at (a) 300, (b) 250, and (c) 220 K, respectively. Horizontal and vertical axes represent frequency of temperature fluctuation and wavelength of static structure factor of center of mass of molecule, respectively. (d) Temperature dependences of the relative contributions of respective shells under constant pressure condition. Red, green blue, pink, aqua, yellow, and gray are the relative contributions of the first, second third, fourth, fifth, sixth, and the other shells.

A growth of correlation is also present under constant volume condition. It should be noted, however, that the growth of correlation length under the condition of constant volume is less significant due to the inhibition of growth of tetrahedral structure caused by the lack of free volume. For example, at 205 K, the relative molecular contribution of the third hydration shell under than condition of constant volume is about 10 %, whereas the same under the constant pressure condition is ~30 %. This result clearly demonstrates that the emergence of spatial correlations among fluctuations is associated with the rapid increase in $C_P$ below ~220 K. Note that the correlation length is greater than nearest neighbor distance and easily reaches about three times that distance near 220 K.

## VI. TEMPERATURE DEPENDENCE OF RELAXATION TIMES: GROWTH NEAR 220 K

Figure 5 displays the temperature dependences of the relaxation times of the HBN dynamics. The significant slow-down of the relaxation time immediately above 220 K is well described by a Vogel-Fulcher-Tammann (VFT) law with the critical temperature of 201 K and the fragility index of 0.59. Below 220 K, however, we found that the temperature dependence of relaxation time can be expressed by another VFT law with the lower critical temperature of 173 K and larger fragility index of 1.25. The second branch can also approximately fitted by an Arrhenius law, but with a large activation energy of ~16.6 kcal/mol. In the convention adopted here, a liquid with a larger fragility index corresponds to a less fragile liquid.

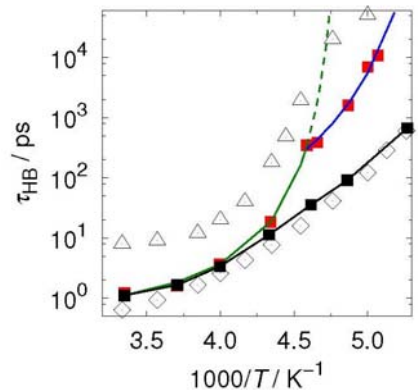

**FIG. 5**: Temperature dependences of the relaxation times of the HBN dynamics under constant pressure (red squares) and constant volume (black squares) conditions. Diamonds and triangles are, respectively, the relaxation times in the self intermediate scattering functions calculated from molecular dynamics simulations at the density $\rho = 1.0$ g/cm$^3$ (ref. 53), and the translational relaxation times obtained from the quasielastic neutron scattering experiment (ref. 54). The red squares above 220 K can be fitted well by a

Vogel-Fulcher-Tammann (VFT) relation with $T_0 = 201$ K (green line) corresponding to a markedly fragile liquid, where as those below 220 K can be fitted by another VFT relation with $T_0 = 173$ K (blue line) corresponding to a weakly fragile liquid.

Furthermore, we found that the isobaric compressibility shows a maximum at almost the same temperature as $C_P$ and $d\rho/dT$ and decreases to a value which is approximately half of the value above 230 K. All the results of the fragility index, stretching parameter given in Table II, and isothermal complexibility indicate that the liquid in lower temperature branch below 220 K is in a less fragile liquid state than the liquid above 230 K. The present result is thus consistent with the so-called fragile-strong transition scenario. It should be noted that the critical temperature, 173 K, could be related to the glass transition temperature proposed by Angell and co-workers.[48-52]

Finally, we would like to compeare the present temperature dependences of the relaxation times of the HBN dynamics with previously reported results. It is noted, first, that the temperature dependence of the relaxation times of the HBN dynamics under constant volume condition is different from that under constant pressure condition; i.e. it hardly shows a noticeable fragile-strong transition found under constant pressure condition. A similar temperature dependence is indeed found in the relaxation time in the self intermediate scattering function calculated from molecular dynamics simulations at the density $\rho = 1.0$ g/cm$^3$ (white diamonds in Fig. 5).[53] As mentioned above, the difference arises from the difference in energy landscape under the two conditions.

Figure 5 also shows the temperature dependce of the translational relaxation time of confined water determined by quasielastic neutron scattering (QENS) experiment.[54] The observed location of the transition temperature and temperature dependence, in particular below the transition temperature, are similar to the present results, although the experimentally determined activation barrier in the second branch is smaller than the present result. Further experimental investigations of the dynamics near and below the transition temperature are required to reveal the properties in the deeply supercooled water.

## VII. CONCLUSIONS

In the present study, we explored, by using long atomistic simulations, the spatio-temporal correlation of temperature fluctuation to understand the anomaly of isobaric specific heat, $C_P$, of supercooled liquid water. First, we analyzed the temperature dependence of isobaric and isochoric specific heats by using molecular dynamics simulations. The calculated $C_P$ rapidly increases as experimentally observed and reaches the maximum at ~220 K and decreases thereafter with further decreasing temperature. On the other hand, such a rapid variation is not observed in the calculated isochoric specific heat, $C_V$.

The time scales and intermolecular motions contributing to the temperature fluctuations have been examined in terms of the complex specific heat. We demonstrate that the slow-down of HBN dynamics with lowering of temperature under constant pressure condition is responsible for the augmentation of specific heat at low frequency. Such a rapid slow-down with decreasing temperature is not seen under constant volume condition. This slow-down is in turn due to the growth of tetrahedral structure arising from the decrease in density, i.e. the increase in local free volume. The complex isobaric specific heat shows not only the slow of the HBN dynamics but also the increase in (temperature fluctuation) weighted density of states, $\hat{C}_P''(\omega)$, of the HBN dynamics compared with the corresponding $\hat{C}_V''(\omega)$.

The present results thus show that the energy landscape under constant pressure condition is different from that under constant volume condition. The difference in energy landscape under the two conditions is also seen in the difference in the temperature dependence of the HBN dynamics.

Perhaps the most important outcome of this study is the result that in addition to the appearance of longer time scale, the length scale contributing to the temperature fluctuations also lengthens as temperature is lowered towards 220 K. We have examined the emergence of length scale correlations in terms of the shell-wise decomposition of the total contribution, and also wavenumber, $k$, dependent temperature fluctuation. At room temperature, more than 75 % of the total fluctuation per molecule arising from HBN dynamics comes from within the first hydration shell. With decreasing temperature, however, the molecular contributions from the distant regions increase; e.g. the contribution extends to the sixth hydration shell at 220 K. The emergence of the correlated region is more significant under constant pressure condition than constant volume condition, because of the growth of tetrahedral structure under the former condition.

We find an interesting crossover from a markedly fragile state to a less fragile state at around 220 K. The liquid below the crossover temperature may be considered to be a weakly fragile liquid based on the fragility index and stretching parameter.

To understand the dynamics in deeply supercooled state and also the questions related to the issue of the presence or absence of a glass transition in water , detail information on the spatial and temporal correlation is essential. Now, it is known that four point correlation functions can probe the length scale of dynamic heterogeneity.[55-59] Systematic analyses using four point correlation functions will be required to understand the growth of spatial correlation and the glass transition, in particular in deeply supercooled water. In addition, multi-time correlation functions have been used to investigate the time scale of heterogeneity.[60-63] We have, recently, found that three-time correlation function is very sensitive to the extent of fragility of the system.[64] Therefore, the analyses based on three-time correlations could shed light on the fragility and temperature dependence of the dynamics in the deeply supercooled water.

Theoretically, we do not seem to have at our disposal a theory that addresses the results and the issues raised here. It would be nice to develop a mode coupling theory approach to temperature fluctuations. Such a work is under progress.

## ACKNOWLEDGMENTS

The present study was supported by Indo-Japan (DST-JSPS) information exchange grant. SS acknowledges the support from JSPS, Grant-in Aid for Scientific Research (No. 22350013) and Grant-in-Aid for Challenging Exploratory Research (No. 23655020), and the supports from Strategic Program for Innovative Research and Computational Material Science Initiative. BB acknowledges the support from JC Bose Fellowship (DST, India). The calculations were partially carried out by using the supercomputers at Research Center for Computational Science in Okazaki.